%% file: main.tex
\documentclass[twocolumn]{fairmeta}

\input{math_commands.tex}

\usepackage{microtype}
\usepackage{graphicx}
\usepackage{subcaption}
\usepackage{booktabs}      

\usepackage[utf8]{inputenc} 
\usepackage[T1]{fontenc}    
\usepackage{url}            
\usepackage{amsfonts}       
\usepackage{nicefrac}       
\usepackage{xcolor}         

\usepackage{amsmath}
\usepackage{amssymb}
\usepackage{mathtools}
\usepackage{amsthm}
\usepackage{float}

\usepackage[textsize=tiny]{todonotes}

\theoremstyle{plain}

\theoremstyle{definition}

\theoremstyle{remark}

\title{Misalignment Between Backpropagation and the Hierarchy of Brain Responses to Images}

\author[1,2]{Joséphine Raugel}
\author[1]{Maximilian Seitzer}
\author[1]{Marc Szafraniec}
\author[1]{Huy V. Vo}
\author[1]{Jérémy Rapin}
\author[1]{Patrick Labatut}
\author[1]{Piotr Bojanowski}
\author[2]{Valentin Wyart}
\author[1]{Jean-Rémi King}

\affiliation[1]{Meta AI}
\affiliation[2]{Ecole Normale Supérieure, PSL University}

\correspondence{\email{josephiner@meta.com, jeanremi@meta.com}}

\abstract{
Backpropagation is the core learning mechanism underlying deep learning.
However, whether and how this algorithm is implemented in the brain remains highly debated.
In particular, while forward activations of pretrained models reliably map onto the cortical hierarchy of visual processing, it is unknown whether backpropagated gradients exhibit a similar correspondence.
Here, we address this question using functional magnetic resonance imaging (fMRI) and magnetoencephalography (MEG) recordings of human brain responses to natural images. For this, we extend standard encoding analyses of forward activations to map backpropagated gradients onto neural data. Focusing on a recent self-supervised vision model (DINOv3) and reproducing results on eight vision models, we find that backpropagated gradients can reliably predict both fMRI and MEG signals, specifically in higher-level visual cortex and for later latencies. However, the spatial and temporal organization of these backpropagated gradients in the brain diverges from the patterns expected under a biologically plausible backpropagation mechanism: specifically, both the order in which gradients
are computed and their spatial organization diverge from the temporal and spatial hierarchies of the human brain. Together, these results suggest that, although deep networks and the brain may share similar representational content, they likely rely on fundamentally different mechanisms to learn those representations.
}

\begin{document}

\maketitle

\section{Introduction}
\begin{figure*}[ht!]
    \includegraphics[width=\textwidth]{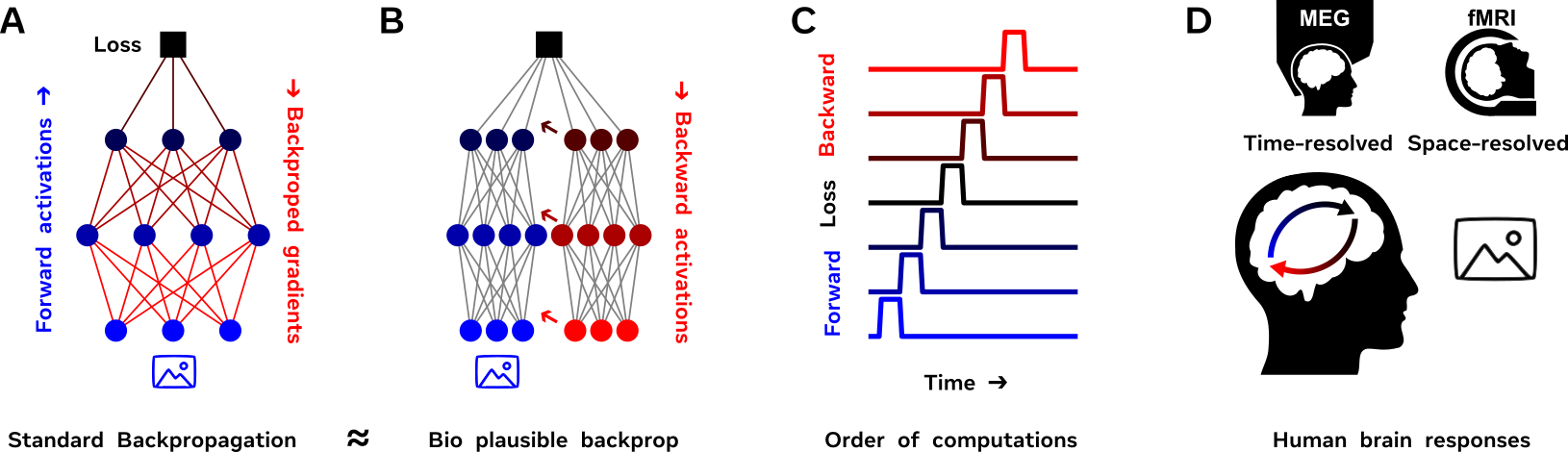}
    \caption{\textbf{Tracking Backpropagation in the Brain.} \textbf{A.} The standard backpropagation algorithm first computes a bottom-up hierarchy of activations, then evaluates the loss, and finally computes a top-down hierarchy of gradients to update network weights. \textbf{B.} A biologically plausible implementation of backpropagation relies on a top-down neural network with a backward stream of activations \citep{lillicrap2020backprop}. This framework predicts that gradient-related information should also be reflected in neural activations. \textbf{C.} Standard and biologically plausible implementations of backpropagation therefore share a similar computational sequence, yielding testable predictions for neuroscience. \textbf{D.} Here, we analyze human brain responses to natural images recorded with MEG and ultra-high-field fMRI (7T), providing high temporal and spatial resolutions. We then evaluate the encoding of forward and backward activations from DINOv3 \citep{dinov3} and other models to characterize the timing and anatomical localization of these representations.
    }
    \label{fig:fig_forward_loss_backprop_square}
\end{figure*}
\textbf{Backpropagation as a learning principle.}
Backpropagation is a core element of deep learning: it computes, in a top-down manner, the gradient of the loss with respect to each synaptic weight in a potentially very deep hierarchical network, enabling loss minimization during training \cite{Rumelhart1986}.
Whether the brain implements backpropagation or a functionally equivalent mechanism remains a central and yet unresolved question in computational neuroscience and artificial intelligence \cite{lillicrap2020backprop, Whittington2017}.

\textbf{Similar visual representations between AIs and the Brain.}
The biological plausibility of this algorithm stands in contrast to that of its "forward" counterpart. Indeed, although deep neural networks are not designed to mimic the brain, their forward activations partially converge toward representations similar to those observed in the mammalian visual system \cite{Yamins2014,KhalighRazavi2014,Guclu2015, cichy2014resolving}. In particular, bottom-up hierarchy of activations in deep networks have repeatedly been found to align with activations of the cortical hierarchy \citep{millet2023realisticmodelspeechprocessing, DinoxBrain_hierarchy_forward_learning, Wang2023}

\textbf{Remaining challenge.}
Empirical comparisons between artificial and biological neural networks have almost exclusively focused on forward activations, leaving open the question of how learning -- and backpropagation in particular -- relates to brain activity. Previous studies have examined neural correlates of prediction errors (\citep{heilbron2022hierarchy} in LLMs). However, to our knowledge, no work has directly tested whether the top-down hierarchy of backpropagated gradients aligns with neural responses.

\textbf{Biological backpropagation.}
While the standard implementation of the backpropagation algorithm requires synaptic symmetry, and is thus biologically implausible, \citet{lillicrap2020backprop} showed that a functionally equivalent but biologically plausible implementation can be computed by approximating backpropagated gradients through a top-down hierarchical neural network. From this perspective, backpropagated gradients can be interpreted as backward activations that convey learning signals across cortical hierarchies.

\textbf{Predictions.}
Under this framework, a biological implementation of backpropagation would give rise to three testable signatures. First, backpropagated gradients from the model should be linearly mappable onto the brain responses elicited by the corresponding stimuli.
Second, mappings based on forward activations should peak earlier than those based on backpropagated gradients.
Third, the spatial and temporal organization of these mappings should reflect a bottom-up progression for forward activations followed by a top-down progression for backpropagated gradients.

\textbf{Approach.}
To test these predictions, we adapt the standard encoding analyses of forward activations to map backpropagated gradients of a SOTA self-supervised vision model (DINOv3) onto the brain response to a large number of natural images.
We combine fMRI \citep{Allen2022} for spatial resolution and MEG \citep{contier2023} for temporal resolution to evaluate whether these signals align with the computational ordering prescribed by backpropagation.
Our results confirm the predictions of a similar forward pass as well as a robust mapping between backpropagated gradients and brain activity, explaining neural variance that could not be explained by forward activations alone, specifically in higher-level visual cortex. However, the temporal and spatial organization of this mapping diverges from the top-down organization of the backpropagation algorithm.

\section{Methods}

\subsection{Encoding}
To evaluate whether, when and where the backpropagated gradient of the loss map onto the brain responses to natural images, we adapt a standard encoding analysis. Following the rationale of \citep{naselaris2011encoding}, encoding provides an interpretable mapping from model features to neural responses.

\textbf{Mapping.} This linear analysis consists of evaluating whether there exists a linear mapping $W\in\mathbb{R}^{d\times m}$ that reliably predicts $m$-dimensional brain activity ($Y\in\mathbb{R}^{n \times m}$) given the $d$-dimensional model activation ($X\in\mathbb{R}^{n\times d}$) in response to $n$ images:
\begin{equation*}
    \underset{W}{\arg\min} \left\{ \| Y - XW \|_2^2 + \lambda \| W \|_2^2 \right\}
\end{equation*}
with $\lambda$ the ridge regularization parameter.
We use linear probes to maintain geometries of compared representations. We use scikit-learn's \texttt{RidgeCV} \citep{pedregosa2011scikit}, 10 logarithmically-spaced regularization $\lambda$ in between $10^0$ and $10^{8}$, and a 5-split cross-validation.

\textbf{Similarity.} We quantify the representational similarity between $X$ and $Y$ by computing, for each split separately an \emph{encoding score} with a Pearson correlation score $R \in[-1, 1]^m$:
\begin{equation*}
    \mathrm{R} =corr(WX_{test}, y_{test})
\end{equation*}
We finally average these encoding scores across splits (and across sensors for MEG signals), leading to 1 encoding score for each subject in each brain location (fMRI) or each time sample (MEG).

\textbf{Statistics.} We compare each voxel's encoding score to a null distribution of 1000 encoding scores obtained with shuffled labels and apply FDR correction to account for the multiple voxels. We retain voxels with FDR-corrected $p < 0.05$. Only these significant voxels are displayed in the brain maps. For time-resolved MEG encoding, significance is assessed at each time point using a one-tailed, one-sample t-test, with FDR correction applied to account for multiple comparisons across time. Time points reaching $p < 0.05$ are indicated as significant with marks * in the figures.

\subsection{Models}
\textbf{DINOv3.} We used DINOv3 (small) as a model \citep{dinov3}. This model is chosen as it is (1) state-of-the-art (2) self-supersived and (3) sufficiently small such that we can analyze the backpropagation gradients with our computational resources. A significant body of literature posits that the brain acts as an active inference engine, seeking to minimize variational free energy rather than merely maximizing external reinforcement. This framework suggests that the human brain relies on predictive coding to learn internal representations of the world through self-supervised error minimization, even in the absence of explicit reward signals \citep{Friston2009}. DINOv3 is selected as a modeling framework because its self-supervised objective may partially align with unsupervised learning principles at play in human brains. This model consists of a student network, optimized to match the representations of a teacher network. The teacher's weights are updated as an exponential moving average (EMA) of the student's weights, providing a slowly evolving, stable target. This setup allows the development of rich, invariant representations without labels. We train DINOv3-S from scratch to analyze the dynamics of Forward and Backward encoding scores across training.

\textbf{Benchmark of eight models.}
We compare eight models spanning four architecture families: self-supervised ViTs (DINOv3-S \citep{dinov3}, ViT-MAE \citep{ViT_MAE}), a supervised ViT (ViT-L/16 \citep{ViT-L}), supervised CNNs (ConvNeXt-L \citep{ConvNext}, ResNet-50 \citep{Resnet}), and vision-language models (OWL-ViT \citep{Owl}, CLIP \citep{Clip}, SigLIP2 \citep{Siglip2}). We compute backpropagated gradients using each model's most natural loss: native reconstruction loss for ViT-MAE, native image-text similarity for VLMs, and CLIP-bridged caption similarity for classification models (ResNet-50, ConvNeXt-L, ViT-L/16). Gradients are collected from the residual block outputs at 10 evenly-spaced blocks throughout the vision encoder.

\textbf{Forward activations and Backpropagated gradients.}
We extract the layer-wise activations of the model for each stimulus. Backpropagated gradients are computed with a standard implementation of the algorithm between the teacher and student networks.
Here, we leverage the image-wise DINOv3 losses, computed for each image seen, and backpropagated across the layers of the model.

\textbf{Encoding of residual Backpropagation.} In order to isolate the specific correlates of backpropagated gradients in the brain, we use a conservative method to remove correlated signals between Backpropagated gradients and Forward representations: we compute the Residual Backpropagation, defined as the encoding score of Forward activations and Backpropagated gradient, subtracted by the encoding score of Forward representation.
\begin{equation*}
\begin{aligned}
R_{\text{Residual Backprop}} = \, &R(\text{Forward} + \text{Backprop}) \\
&- R(\text{Forward})
\end{aligned}
\end{equation*}

\textbf{Layer-wise extraction.} We analyze 12 layers, indexed by their normalized depth $k \in [0, 1]$, where $k=0$ and $k=1$ correspond to the earliest and latest stages of the model, respectively. Specifically, we extract the activations from the projection layer (W3) of the MLP within each of the 12 transformer blocks.
For each layer, we get the Forward $X_{forward}\in\mathbb{R}^{n\times d}$ and Backward signals $X_{backprop}\in\mathbb{R}^{n\times d'}$ and evaluate their linear mappings to the brain activity $Y\in\mathbb{R}^{n\times m}$ as the encoding score.

\textbf{Random Projection and PCA.} Backward signals have a high dimensionality of 4e{6}. To be able to store and analyze these signals within our computational resources, we perform projection of these signals onto a random matrix of dimension 800. Following the rationale of \citep{Halko2011} this method approximates to a subspace PCA in high dimensionality.
To mitigate the issue of heterogeneous vector sizes, we then transform these Forward $X_{forward}\in\mathbb{R}^{n\times d}$ and Backwards vectors $X_{backprop}\in\mathbb{R}^{n\times d'}$ with a Principal Component Analysis (n=300) with scikit-learn \citep{pedregosa2011scikit} prior to encoding analyses. This preprocessing results in Forward $X_{forward}\in\mathbb{R}^{n\times300}$ and Backward vectors $X_{backprop}\in\mathbb{R}^{n\times300}$ of identical dimensions. To compute Residual Backpropagation, we compare the concatenation of $X_{forward}\in\mathbb{R}^{n\times150}$ and $X_{backprop}\in\mathbb{R}^{n\times150}$ to $X_{forward}\in\mathbb{R}^{n\times300}$, allowing to compare features in the same dimensionality. We conduct a grid search across Projection dimensions and PCA dimensions to ensure the robustness of our results, Supp. Fig. \ref{fig:pca_projection_robustness}.

\subsection{Datasets}
\textbf{Magnetoencephalography (MEG).} We use the THINGS-MEG dataset \citep{contier2023}, which consists of MEG recordings from four healthy participants viewing 22,500 naturalistic images, representing a total of 1,800 object concepts. Images were presented during 1.5\,s, while participants maintained fixation. To limit the impact of noise we apply a bandpass filter between 0.1 and 20\,Hz, down-sample the signal at 30\,Hz, time-lock the brain responses to individual images, and epoch the corresponding neural data between -0.5\,s and +3\,s relative to image onset using MNE-Python \citep{gramfort2013meg} and NeuroAI~\citep{king2026neuralset}. Finally, we z-score MEG signals across images, for each MEG channel and each time point independently.

\textbf{Functional Magnetic Resonance Imaging (fMRI).} We leverage the Natural Scenes Dataset \citep{Allen2022}, a 7 tesla fMRI dataset which consists of recordings from eight subjects, each observing a total of 10 000 natural scenes during 4 seconds each, while performing a continuous recognition task. We encode fMRI beta responses time-locked to image onset, on the fsaverage surface. Across analyses, we study only voxels significantly encoded.

\textbf{Best Encoded Regions of interest (ROIs).} We select the best-encoded ROIs, defined as those whose mean Forward encoding score falls within the top tercile ($33\%$) of all mean ROI scores.

\section{Results}
\begin{figure*}[t!]
    \centering
    \includegraphics[width=\textwidth]{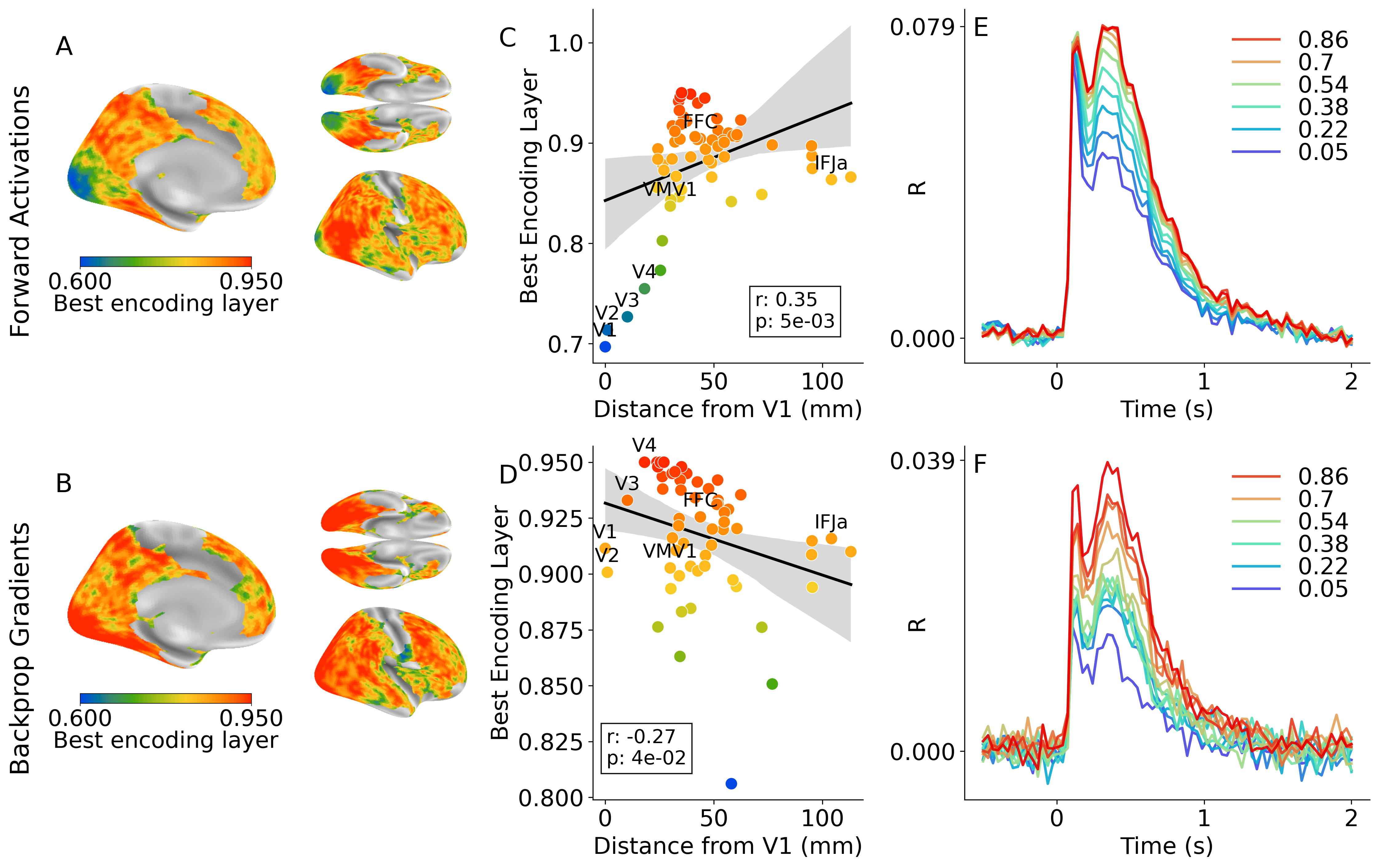}
    \caption{\textbf{Dissociation between forward and backward hierarchical alignment with the brain.}
    Panels A--B: fMRI cortical surface maps of the best encoding layer for Forward Activations and Backpropagated Gradients - approximated through PCA - at the end of training from scratch (at iteration 500k). Surface-smoothed (3 iterations, geodesic neighbor averaging, weight = 0.3). Panels C--D: Pearson correlation between ROI distance from V1 and best encoding layer, for the top third best encoded ROIs. Panels E--F: Time-resolved MEG encoding curves for each of the 12 studied layers.}
    \label{fig:fig_backprop_hierarchy}
\end{figure*}

\paragraph{Anatomical organization.}
We aim to identify whether forward and backward passes of a self-supervised vision model such as DINOv3 align with the anatomical organization of the human brain.\\
Our results indicate that in the forward pass, early visual areas are best predicted by shallow layers, while associative cortex is best predicted by deeper layers (Fig. \ref{fig:fig_backprop_hierarchy}A), reflecting a bottom-up correspondence with cortical hierarchy. We find a significant positive correlation (Pearson $r = 0.35$, $p = 5\text{e-}3$, Spearman $\rho = 0.27$, $p = 3\text{e-}2$) between the distances of best-encoded ROIs from V1 and the layer that encode them best (Fig. \ref{fig:fig_backprop_hierarchy}C).
These results replicate previous findings in the literature regarding a cohort of self-supervised and supervised models \citep{DinoxBrain_hierarchy_forward_learning, Wang2023}.

Conversely, in the backward pass, early sensory areas are best aligned with deep gradients, whereas associative regions align with shallower gradients. (Fig. \ref{fig:fig_backprop_hierarchy}B). We find a negative spatial correlation (Pearson $r = -0.27$, $p = 4\text{e-}2$, Spearman $\rho = -0.34$, $p = 8\text{e-}3$) between the distances from V1 of the selected ROIs, and the layer that encode them best (Fig. \ref{fig:fig_backprop_hierarchy}F). This negative hierarchical trend indicates that gradient-based features do not follow the layer-to-cortical-area mapping observed in forward representations, and may even tend to reverse it.

Overall, the forward pass of DINOv3 aligns with the bottom-up anatomical organization of the human brain, whereas the backward pass does not align with the top-down organization of backpropagated gradients along layers of DINOv3.

\paragraph{Temporal organization.}
We aim to further compare the temporal organization of the activations in the human brain with the hierarchy of DINOv3, regarding feedforward and backpropagated signals.
We first assess the temporal alignment of forward activations of DINOv3 with brain responses.

The peaking times of encoding scores correspond to the order of their respective encoding layers:
all layers peak around the same encoding score (R= 0.079), but shallower layers peak earlier, whereas deeper layers align later with brain responses (Fig. \ref{fig:fig_backprop_hierarchy}E).\\
Conversely, when considering backpropagated signals, peaks of alignment are not sequentially organized according to layer depth: all layers peak at the same time while a gradient in the peaking score is observed, with earlier layers peaking lower than later layers (Fig. \ref{fig:fig_backprop_hierarchy}F).
This suggests that while feedforward activations in DINOv3 reflect a hierarchical temporal structure akin to that observed in the brain, backpropagated signals do not exhibit such a clear temporal hierarchy. Notably, backpropagated gradients across all layers peak simultaneously with the final layer of the forward pass. This lack of temporal delay contradicts biologically plausible implementations of backpropagation, which would predict backward signals to peak sequentially following the forward pass, Fig. \ref{fig:fig_forward_loss_backprop_square}.

\paragraph{Linear mapping of forward activations and backpropagated gradients is consistent across architectures.}
\begin{figure*}[t!]
    \centering
    \includegraphics[width=\textwidth]{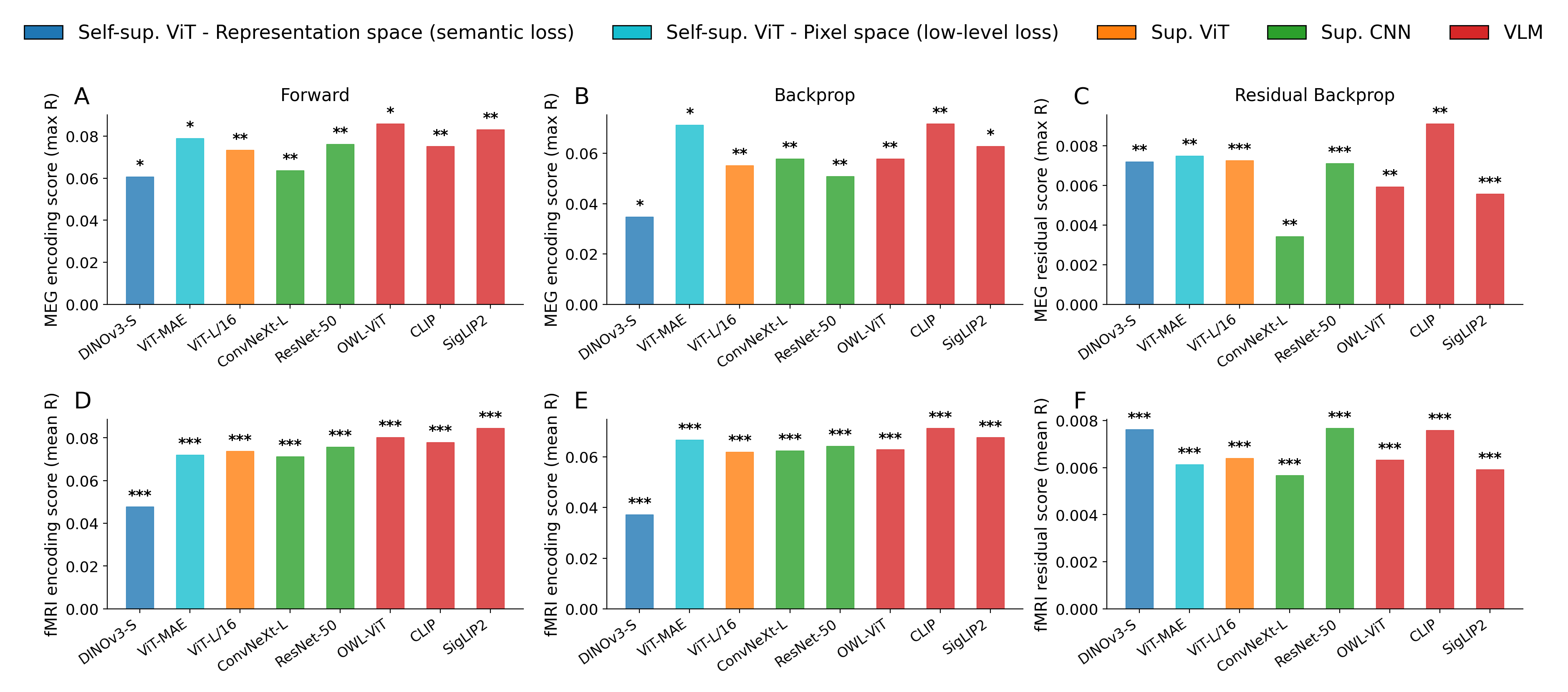}
    \caption{\textbf{Generalized neural alignment of forward and backward activations across diverse vision architectures.} Encoding scores of the last layer for eight vision models across Forward, Backprop, and Residual Backprop conditions. Top row: MEG encoding scores (maximum Pearson $r$ across time of the mean-across-subjects temporal curve). Bottom row: fMRI encoding scores (mean Pearson $r$ across significant voxels). Models are grouped by category: self-supervised ViTs trained with semantic loss (DINOv3-S) or pixel-level reconstruction loss (ViT-MAE), supervised ViT (ViT-L/16), supervised CNNs (ConvNeXt-L, ResNet-50), and vision-language models (OWL-ViT, CLIP, SigLIP2). Significancy: * $p < .05$, ** $p < .01$, *** $p < .001$.}
    \label{fig:encoding_scores_barplot}
\end{figure*}

\begin{figure*}[t!]
    \centering
    \includegraphics[width=\textwidth]{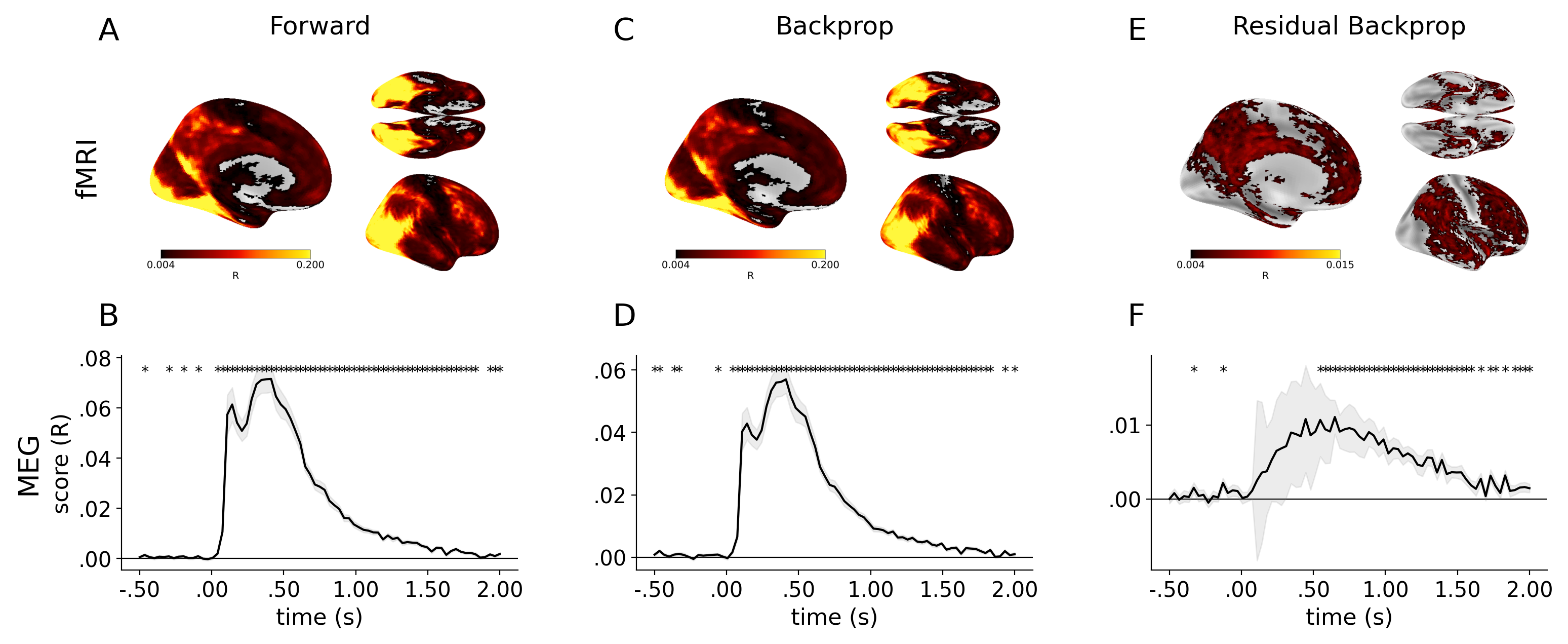}
    \caption{\textbf{Neural alignment over time and space with forward, backward and residual backward activations, averaged across models}. Top row: fMRI cortical surface maps showing the mean encoding score (Pearson $r$) across models for Forward (A), Backward (C), and Residual Backward (E) activations. Only significant voxels are plotted. Bottom row: MEG temporal encoding curves showing the mean score (black line) $\pm$ SEM (gray shading) across all subjects and models, for Forward (B), Backward (D), and Residual Backward (F) activations. Asterisks denote timepoints with significant encoding (FDR-corrected $p < 0.05$).}
    \label{fig:avg_across_models_fdr}
\end{figure*}

Backpropagated gradients from the self-supervised DINOv3-S model show significant alignment with both hemodynamic (fMRI) and electrophysiological (MEG) brain responses, Fig. \ref{fig:fig_backprop_hierarchy}, \ref{fig:encoding_scores_barplot}. How does this mapping evolve across diverse architectures and learning objectives?\\
We compare eight models spanning four families of architectures and learning objectives: self-supervised ViTs (DINOv3-S, ViT-MAE), a supervised ViT (ViT-L/16), supervised CNNs (ConvNeXt-L, ResNet-50), and vision-language models (OWL-ViT, CLIP, SigLIP2).
We investigate encoding scores of both forward and backward passes across processing time (MEG) and cortical space (fMRI).

First, results confirm that forward activations of these models can be mapped onto these brain recordings (Fig. \ref{fig:encoding_scores_barplot}A, D, \ref{fig:avg_across_models_fdr}A, B). Forward activations significantly map onto brain responses in the visual system in fMRI, and shortly after image onset in MEG. This phenomenon was previously observed in the literature \citep{schrimpf2018brain, dicarlo, Zhuang2021}.

We additionally find that backpropagated gradients of each of these models also significantly map onto the human brain recordings, Fig. \ref{fig:encoding_scores_barplot}B, E. Similar regions of the visual system and processing latencies are engaged by both Forward and Backprop activity, indicating that Backprop specifically encodes the synaptic updating occurring in these visual regions, Fig. \ref{fig:avg_across_models_fdr}C, D.

Finally, we compute the Residual Backpropagation, a conservative measure which isolates gradient-specific neural variance beyond Forward activations. Residual Backpropagation scores are obtained from backpropagated features after removing the variance explained by Forward activations. Residual Backpropagation still significantly predicts brain activity, demonstrating that backpropagated gradients capture variance in neural responses that is not accounted for by Forward activations alone, Fig. \ref{fig:encoding_scores_barplot}C, F.

Interestingly, Residual Backpropagation encodes most accurately representations of the late visual cortex, Fig. \ref{fig:avg_across_models_fdr}E.
This observation is corroborated by MEG results, where residual backpropagation demonstrates its highest encoding performance during later latencies, between 0.5s and 1.7s after image onset, Fig. \ref{fig:avg_across_models_fdr}F. Taken together, these results indicate that residual backpropagation primarily align with higher-level representations in the late visual cortex.

Overall, these results show that backpropagated gradients significantly map onto brain activity during passive viewing of images and explain neural responses beyond what is captured by Forward activations. This finding suggests that the brain might implement learning mechanisms that share similarities with backpropagation, such as predictive coding \citep{lillicrap2020backprop, Bastos2012}.

Additionally, we reproduce the temporal and spatial gradients observed in Fig. \ref{fig:fig_backprop_hierarchy} for Forward and Backward activations, combining these eight models, Supp. Fig. \ref{fig:combined_8models}.

\paragraph{Training Dynamics of Forward Activations, Backpropagated Gradients and Residual Backpropagation.}
\begin{figure*}[t!]
    \includegraphics[width=\textwidth]{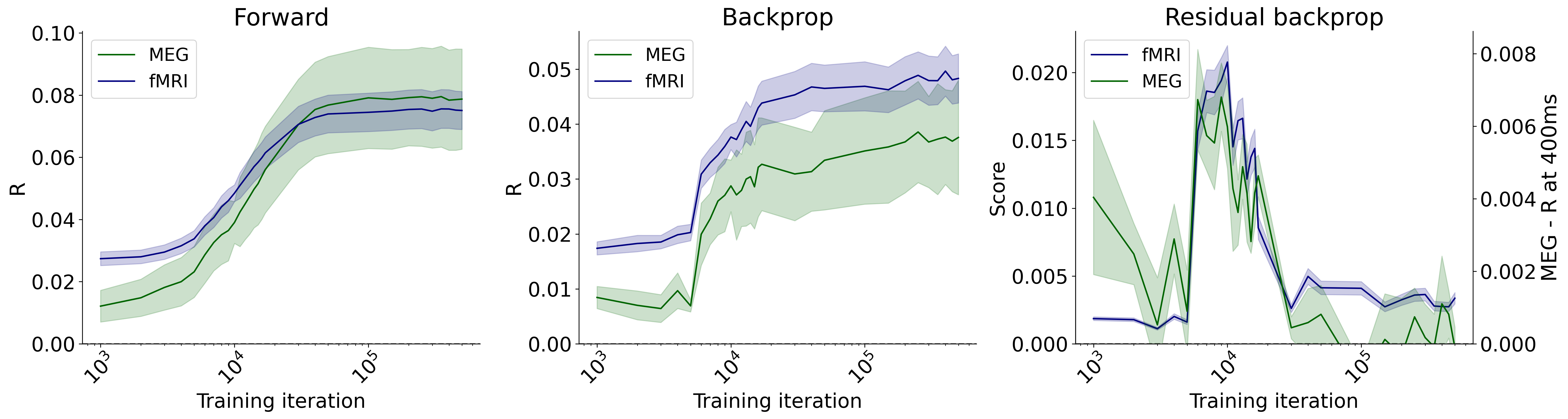}
    \caption{\textbf{Evolution of Forward, Backpropagation, and Residual Signals Across Model Training.}
    Results are shown for Forward (left), Backpropagation (middle), and Residual Backpropagation (right) encoding scores for the last layer.
    Both MEG (green) and fMRI (blue) scores exhibit synchronized trajectories across training iterations.
    Errorbars indicate the standard error across subjects for each modality.}
    \label{fig:across_training}
\end{figure*}

\begin{figure*}[ht!]
\centering
    \includegraphics[width=0.9\textwidth]{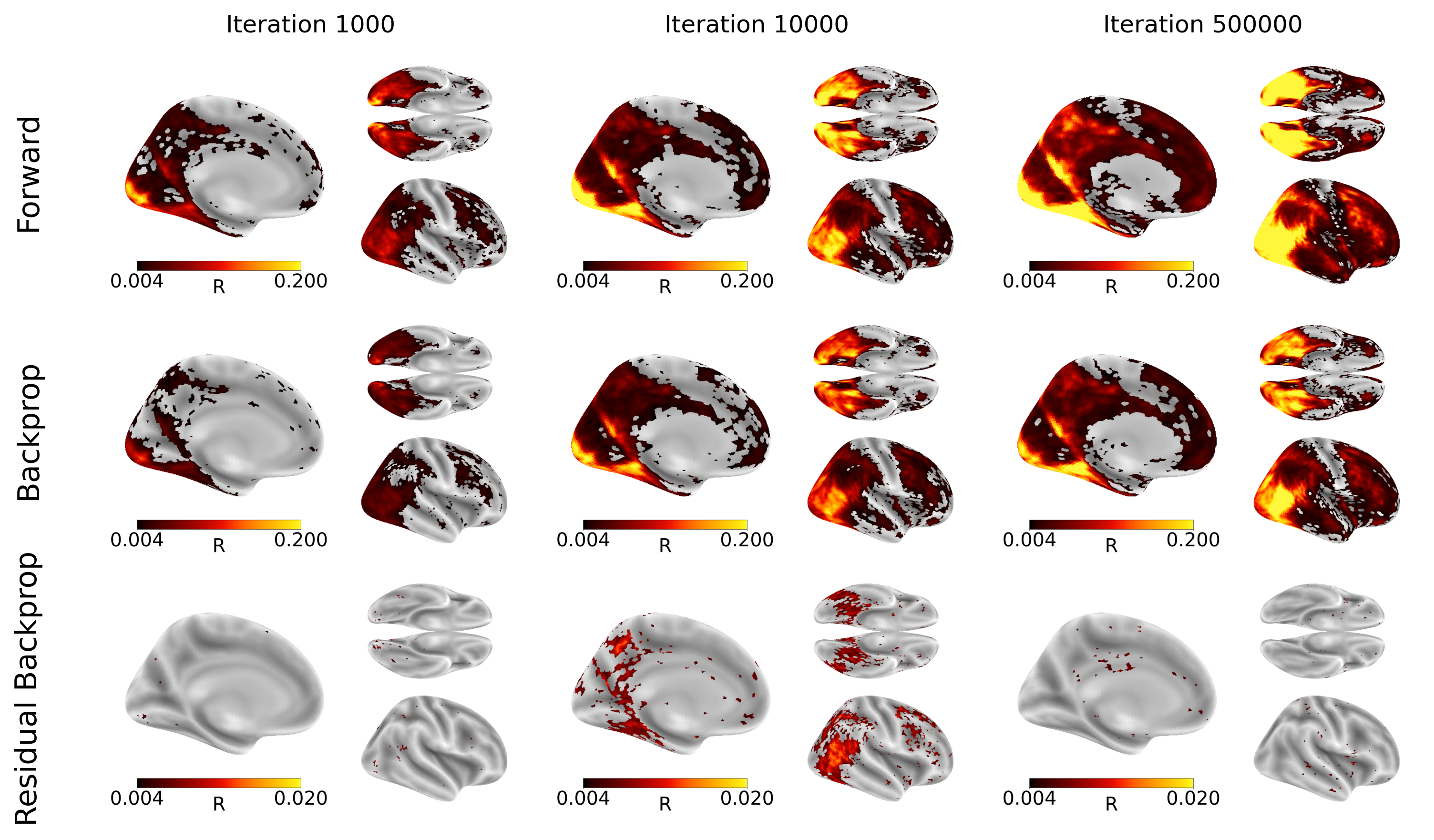}
       \caption{\textbf{Spatial Distribution of Forward, Backpropagation, and Residual Backpropagation across training.} FMRI encoding scores for Forward Activations (up), Backpropagated Gradients (middle), and Residual Backpropagated Gradients (Bottom) at three training stages for the deeper layer. Only significant voxels ($p<0.05$) are shown.}
    \label{fig:fmri_layer95}
\end{figure*}

\begin{figure*}[ht!]
\centering
    \includegraphics[width=0.9\textwidth]{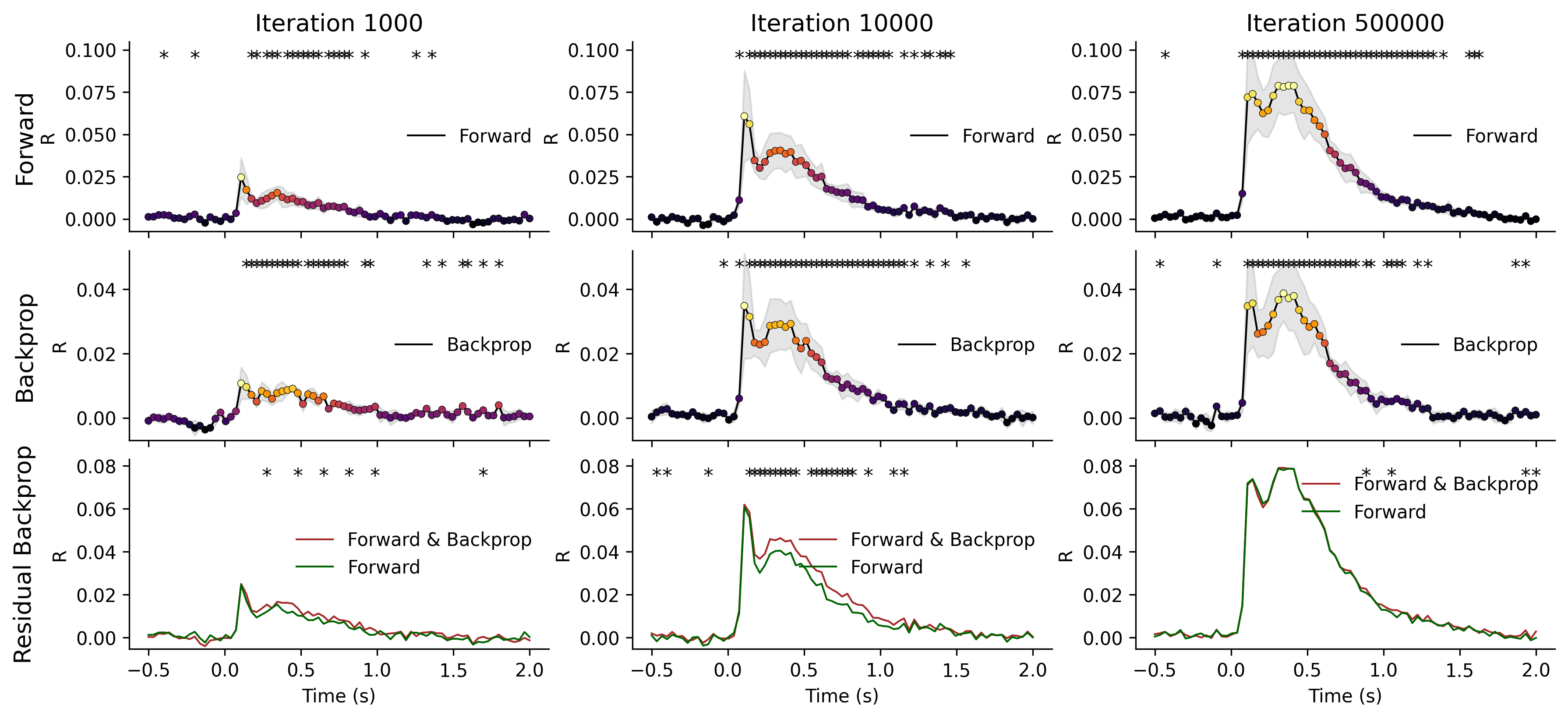}
    \caption{\textbf{Temporal Distribution of Forward, Backpropagation, and Residual Signals across training.} MEG encoding scores for Forward Activations (up), Backpropagated Gradients (middle), and Residual Backpropagated Gradients (Bottom) at three training stages for the deeper layer. Shaded areas represent the standard error across subjects. Asterisks denote timepoints with significant encoding ($p<0.05$).}
    \label{fig:MEG_layer95}
\end{figure*}
We wonder how Forward and Backpropagated signals align with the human brain across model training. To study this question, we train from scratch DINOv3-S on naturalistic data. We chose DINOv3 under the assumption that its SSL objective may be the closest approximate -- among the eight studied models -- of the brain's capacity for representation learning through passive, unsupervised environmental exposure.

Encoding scores from forward and backpropagated signals increase logarithmically throughout training from scratch, exhibiting consistent trajectories across both MEG and fMRI modalities, Fig. \ref{fig:across_training}. Both Forward and Backpropagation scores expand across training towards the higher-level regions of the cortex, and towards later latencies, Fig. \ref{fig:fmri_layer95}, \ref{fig:MEG_layer95}.

In contrast, residual backpropagation scores follow a distinct non-monotonic trajectory: they initially increase, reach a distinct peak at 10k iterations, and subsequently decrease. This pattern is consistent across both neural recording methods fMRI and MEG, demonstrating a synchronized evolution across both temporal and spatial scales. At peak time (10k iterations), Residual Backpropagation significantly encodes late-latency neural activations, and associative brain regions in the late visual cortex, involved in higher-level processing -- ventromedial and lateral-occipitotemporal cortex.
This transient peak suggests the emergence a learning signal that intensifies across training as gradients align with meaningful updates, then subsides as the DINOv3 teacher and student networks converge.

\section{Discussion}

\textbf{Summary of findings.}
In this study, we investigated the alignment between DINOv3 \citep{dinov3}, a SOTA self-supervised vision transformer, and the human brain response to natural images.
We found that backpropagated gradients from DINOv3 can be linearly mapped onto fMRI and MEG responses to natural images, indicating that gradient-related signals are reflected in brain activity. These results were replicated across 8 diverse vision models.\\
Importantly, even after removing the contribution of forward activations, backpropagated gradients still explained a significant portion of neural responses. This indicates that gradients capture additional variance in brain activity beyond forward representations, specifically for later latencies and in higher-level ROIs -- e.g. multisensory integration, memory, semantic processing. This alignment may reflect intensive synaptic updating regarding high-level representation.\\
However, the temporal and spatial organization of this mapping does not match predictions from a biologically plausible implementation of backpropagation. First, backward encoding does not peak later than forward encoding. Second, contrary to what can be observed with the forward activations, the expected top-down hierarchy of backpropagated gradients does not match the anatomical and temporal organization of brain responses.

\textbf{Relation to biological backpropagation theories.}
A substantial body of prior work has examined how backpropagation, or functionally equivalent learning rules, could in principle be implemented in biological neural networks. Much of this effort has focused on resolving the weight transport problem -- the requirement that feedback connections mirror feedforward weights -- which is not supported by cortical circuitry. Proposed solutions, including random feedback alignment and segregated feedforward and feedback pathways, demonstrate that precise synaptic symmetry is not strictly necessary for gradient-based learning \citep{lillicrap2020backprop}. Importantly, these variants preserve the core computational structure of backpropagation: error signals are computed at higher levels and propagated downward to guide synaptic updates at earlier stages. Our findings do not contradict the theoretical feasibility of such learning mechanisms. Rather, they provide empirical constraints, suggesting that if a gradient-based learning operates in the brain, it does not manifest with the spatial and temporal signatures predicted by backpropagation.

\textbf{Implications for learning mechanisms in the brain.}
The observed dissociation between representational similarity and learning dynamics suggests that shared representations between deep networks and the brain do not necessarily imply shared learning algorithms. While forward activations in deep models reliably mirror cortical hierarchies, the organization of gradient-related signals does not follow the canonical bottom-up / top-down sequence implied by backpropagation. This raises the possibility that the brain relies on alternative learning mechanisms that give rise to similar representational structures without explicitly computing or propagating error gradients. Candidate frameworks include local Hebbian or contrastive learning rules \citep{zenke2017temporal}, predictive coding architectures \citep{Bastos2012,oord2018representation}, and other biologically grounded credit-assignment strategies. Discriminating among these possibilities will require models that make explicit, testable predictions about both neural activity and learning dynamics.

\textbf{Modeling limitations.}
While representational mapping was evaluated on a broad set of eight models, the in-depth training dynamics analysis was restricted to a single SSL model, DINOv3-S, chosen for its strong performance and well-established correspondence with visual cortical representations. Focusing on a SSL model ensures that the learning mechanism could, in principle, be biologically plausible, as it does not require additional information, such as the image labels. However, learning mechanisms of DINOv3 such as the use of EMA and the batch-based nature of backpropagation may not directly map onto human learning strategies. Additionally, unlike the brain, DINOv3 operates on static images and thus lacks explicit temporal dynamics or recurrent processing, limiting its ability to capturing learning signals that unfold over space rather than over time. Future research should aim to refine these models to better mimic biological learning processes and further elucidate the computational principles underlying brain function.
Future work should also extend this approach to a broader range of architectures, including recurrent or predictive models, and video-based systems that may use backpropagation with a different type of prediction objective (e.g. predicting the next frame).

\textbf{Measurement limitations.}
Our conclusions are also constrained by the spatial and temporal resolution of non-invasive human neuroimaging. While fMRI and MEG provide complementary coverage of spatial and temporal dynamics, they cannot directly resolve laminar-specific activity or synapse-level learning signals. Invasive recordings or animal models with controlled learning paradigms may be necessary to more directly probe the neural implementation of credit assignment.

\textbf{Impact statement.}
The human brain remains extraordinarily more data -- and energy -- efficient than recent deep networks. Understanding its learning mechanisms is thus a key goal for both neuroscience and AI. The divergence between the human brain and the backward parts of a SOTA self-supervised model \citep{dinov3} suggests that there exists a different -- potentially more efficient -- strategy to update a large number of synaptic weights. Understanding this biological algorithm could inspire novel, more efficient, robust, and generalizable approaches to machine learning, and -- perhaps more importantly -- this approach could illuminate the biological principles of intelligence.

\newpage
\bibliographystyle{assets/plainnat}
\bibliography{main_bib}



\newcommand{\beginsupplement}{
    \setcounter{table}{0}
    \renewcommand{\thetable}{S\arabic{table}}%
    \setcounter{figure}{0}
    \renewcommand{\thefigure}{S\arabic{figure}}%
    \setcounter{equation}{0}
    \renewcommand{\theequation}{S\arabic{equation}}%
}
\beginsupplement

\appendix

\section*{Acknowledgements} 
The authors wish to thank the Max Planck Institute of Human Cognitive and Brain Sciences in Leipzig, Germany, particularly Professor Dr. Martin Hebart and Dr. Olvier Contier, for providing catch images to explore further the \cite{contier2023} dataset.

\section*{Supplementary Materials}
\begin{figure*}[ht!]
    \centering
    \includegraphics[width=\textwidth]{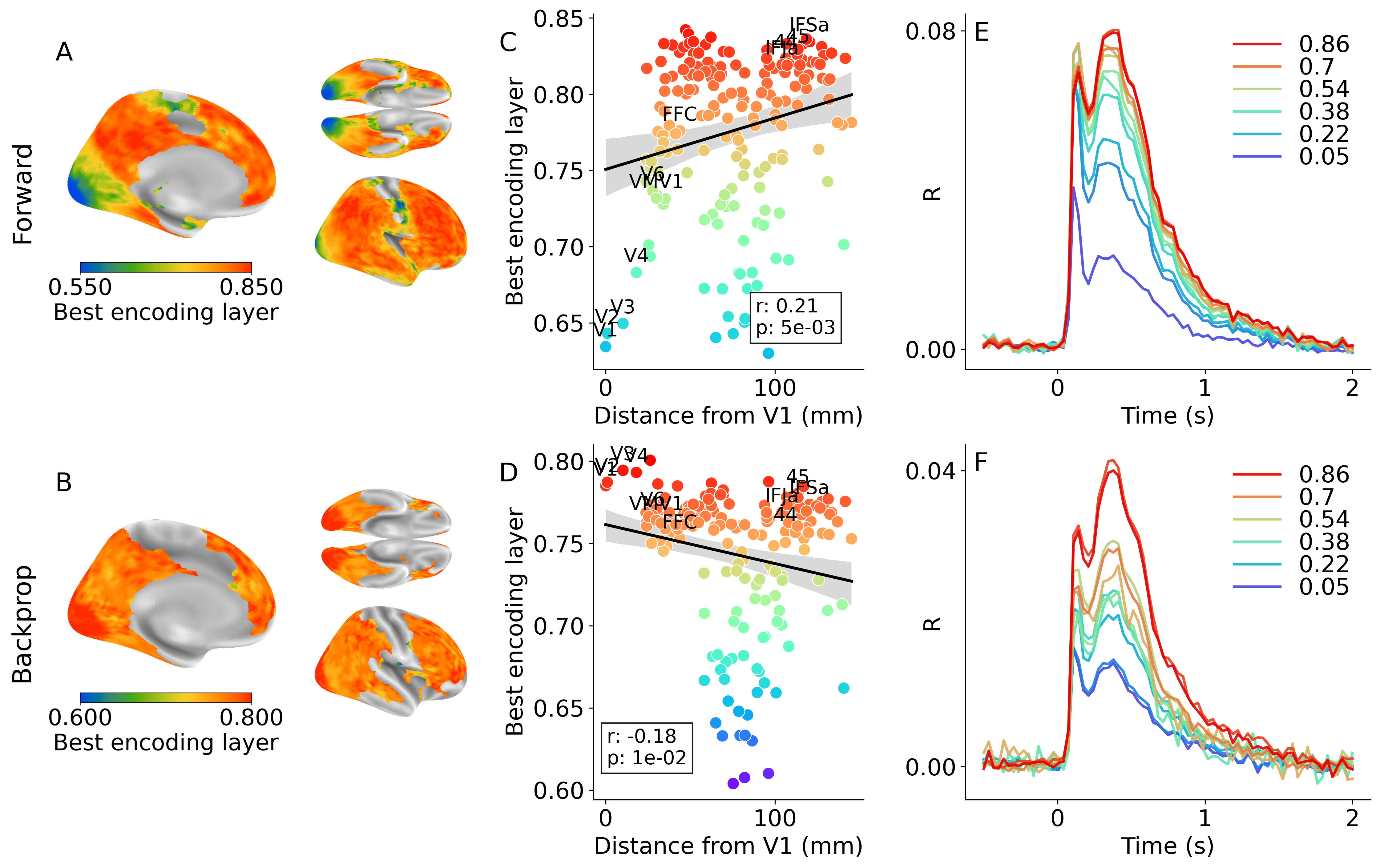}
    \caption{DINOv3-S encoding across all layers, combined across a combination of eight vision models spanning diverse architectures. Panels A--B: fMRI cortical surface maps of the best encoding layer (averaged across the three fMRI datasets) for Forward and Backprop representations. Surface-smoothed (3 iterations, geodesic neighbor averaging, weight = 0.3). Only significant voxels plotted (FDR-corrected p $<$ 0.05 among the 8 model scores). Panels C--D: Pearson correlation between each ROI's distance from V1 and its best encoding layer. Panels E--F: MEG/EEG temporal encoding curves for each layer (averaged across subjects from all three MEG/EEG datasets), for Forward and Backprop.}
    \label{fig:combined_8models}
\end{figure*}

\begin{figure*}[ht!]
    \includegraphics[width=\linewidth]{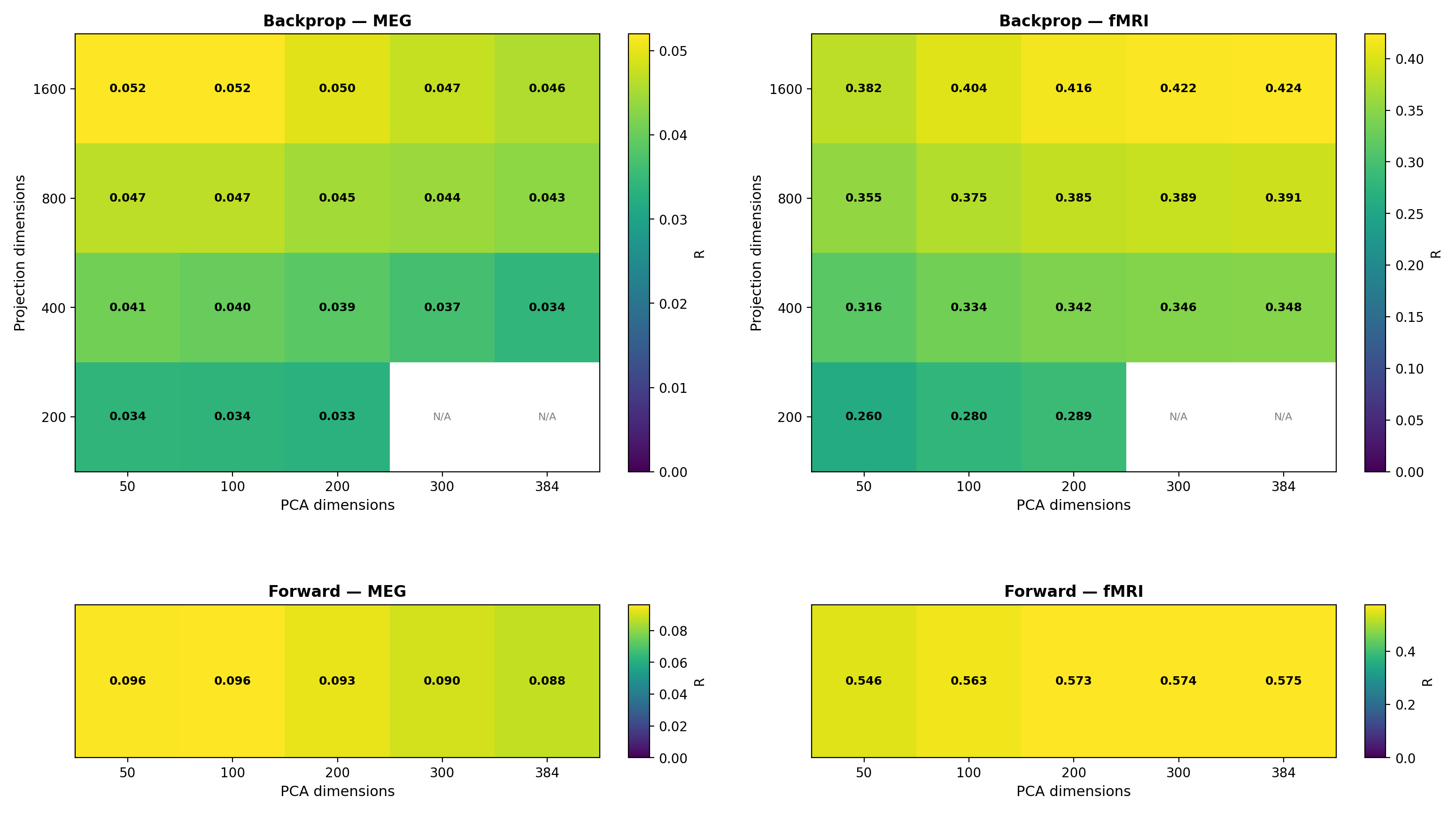}
    \caption{\textbf{Robustness to PCA dimensionality and projection dimension.} (Top row) Backprop encoding score as a function of PCA dimensions (x-axis) and random projection dimensions (y-axis) for MEG (left) and fMRI (right). (Bottom row) Forward encoding score vs. PCA dimensions (no random projection applied).}
    \label{fig:pca_projection_robustness}
\end{figure*}

\end{document}

%% file: math_commands.tex
\usepackage{amsmath,amsfonts,bm}

\def\eqref#1{equation~\ref{#1}}

\def\1{\bm{1}}

\DeclareMathAlphabet{\mathsfit}{\encodingdefault}{\sfdefault}{m}{sl}
\SetMathAlphabet{\mathsfit}{bold}{\encodingdefault}{\sfdefault}{bx}{n}














%% file: main.bbl
\begin{thebibliography}{33}
\providecommand{\natexlab}[1]{#1}
\providecommand{\url}[1]{\texttt{#1}}
\expandafter\ifx\csname urlstyle\endcsname\relax
  \providecommand{\doi}[1]{doi: #1}\else
  \providecommand{\doi}{doi: \begingroup \urlstyle{rm}\Url}\fi

\bibitem[Allen et~al.(2022)Allen, St-Yves, Wu, Breedlove, Prince, Dowdle, Nau, Caron, Pestilli, Charest, Hutchinson, Naselaris, and Kay]{Allen2022}
Emily~J. Allen, Ghislain St-Yves, Yihan Wu, Jesse~L. Breedlove, Jacob~S. Prince, Logan~T. Dowdle, Matthias Nau, Brad Caron, Franco Pestilli, Ian Charest, J.~Benjamin Hutchinson, Thomas Naselaris, and Kendrick Kay.
\newblock A massive 7t fmri dataset to bridge cognitive neuroscience and artificial intelligence.
\newblock \emph{Nature Neuroscience}, 25\penalty0 (1):\penalty0 116–126, January 2022.
\newblock ISSN 1546-1726.
\newblock \doi{10.1038/s41593-021-00962-x}.
\newblock \url{http://dx.doi.org/10.1038/s41593-021-00962-x}.

\bibitem[Bastos et~al.(2012)Bastos, Usrey, Adams, Mangun, Fries, and Friston]{Bastos2012}
Andre~M. Bastos, W.~Martin Usrey, Rick~A. Adams, George~R. Mangun, Pascal Fries, and Karl~J. Friston.
\newblock Canonical microcircuits for predictive coding.
\newblock \emph{Neuron}, 76\penalty0 (4):\penalty0 695–711, November 2012.
\newblock ISSN 0896-6273.
\newblock \doi{10.1016/j.neuron.2012.10.038}.
\newblock \url{http://dx.doi.org/10.1016/j.neuron.2012.10.038}.

\bibitem[Cichy et~al.(2014)Cichy, Pantazis, and Oliva]{cichy2014resolving}
Radoslaw~Martin Cichy, Dimitrios Pantazis, and Aude Oliva.
\newblock Resolving human object recognition in space and time.
\newblock \emph{Nature neuroscience}, 17\penalty0 (3):\penalty0 455--462, 2014.

\bibitem[DiCarlo et~al.(2012)DiCarlo, Zoccolan, and Rust]{dicarlo}
James~J DiCarlo, Davide Zoccolan, and Nicole~C Rust.
\newblock How does the brain solve visual object recognition?
\newblock \emph{Neuron}, 73\penalty0 (3):\penalty0 415--434, 2012.

\bibitem[Dosovitskiy et~al.(2020)Dosovitskiy, Beyer, Kolesnikov, Weissenborn, Zhai, Unterthiner, Dehghani, Minderer, Heigold, Gelly, Uszkoreit, and Houlsby]{ViT-L}
Alexey Dosovitskiy, Lucas Beyer, Alexander Kolesnikov, Dirk Weissenborn, Xiaohua Zhai, Thomas Unterthiner, Mostafa Dehghani, Matthias Minderer, Georg Heigold, Sylvain Gelly, Jakob Uszkoreit, and Neil Houlsby.
\newblock An image is worth 16x16 words: Transformers for image recognition at scale, 2020.
\newblock \url{https://arxiv.org/abs/2010.11929}.

\bibitem[Friston and Kiebel(2009)]{Friston2009}
Karl Friston and Stefan Kiebel.
\newblock Predictive coding under the free-energy principle.
\newblock \emph{Philosophical Transactions of the Royal Society B: Biological Sciences}, 364\penalty0 (1521):\penalty0 1211–1221, May 2009.
\newblock ISSN 1471-2970.
\newblock \doi{10.1098/rstb.2008.0300}.
\newblock \url{http://dx.doi.org/10.1098/rstb.2008.0300}.

\bibitem[Gramfort et~al.(2013)Gramfort, Luessi, Larson, Engemann, Strohmeier, Brodbeck, Goj, Jas, Brooks, Parkkonen, et~al.]{gramfort2013meg}
Alexandre Gramfort, Martin Luessi, Eric Larson, Denis~A Engemann, Daniel Strohmeier, Christian Brodbeck, Roman Goj, Mainak Jas, Teon Brooks, Lauri Parkkonen, et~al.
\newblock Meg and eeg data analysis with mne-python.
\newblock \emph{Frontiers in Neuroinformatics}, 7:\penalty0 267, 2013.

\bibitem[Guclu and van Gerven(2015)]{Guclu2015}
U.~Guclu and M.~A.~J. van Gerven.
\newblock Deep neural networks reveal a gradient in the complexity of neural representations across the ventral stream.
\newblock \emph{Journal of Neuroscience}, 35\penalty0 (27):\penalty0 10005–10014, July 2015.
\newblock ISSN 1529-2401.
\newblock \doi{10.1523/jneurosci.5023-14.2015}.
\newblock \url{http://dx.doi.org/10.1523/JNEUROSCI.5023-14.2015}.

\bibitem[Halko et~al.(2011)Halko, Martinsson, and Tropp]{Halko2011}
N.~Halko, P.~G. Martinsson, and J.~A. Tropp.
\newblock Finding structure with randomness: Probabilistic algorithms for constructing approximate matrix decompositions.
\newblock \emph{SIAM Review}, 53\penalty0 (2):\penalty0 217–288, January 2011.
\newblock ISSN 1095-7200.
\newblock \doi{10.1137/090771806}.
\newblock \url{http://dx.doi.org/10.1137/090771806}.

\bibitem[He et~al.(2016)He, Zhang, Ren, and Sun]{Resnet}
Kaiming He, Xiangyu Zhang, Shaoqing Ren, and Jian Sun.
\newblock Deep residual learning for image recognition.
\newblock In \emph{2016 IEEE Conference on Computer Vision and Pattern Recognition (CVPR)}, page 770–778. IEEE, 2016.
\newblock \doi{10.1109/cvpr.2016.90}.
\newblock \url{http://dx.doi.org/10.1109/CVPR.2016.90}.

\bibitem[He et~al.(2022)He, Chen, Xie, Li, Dollar, and Girshick]{ViT_MAE}
Kaiming He, Xinlei Chen, Saining Xie, Yanghao Li, Piotr Dollar, and Ross Girshick.
\newblock Masked autoencoders are scalable vision learners.
\newblock In \emph{2022 IEEE/CVF Conference on Computer Vision and Pattern Recognition (CVPR)}, page 15979–15988. IEEE, 2022.
\newblock \doi{10.1109/cvpr52688.2022.01553}.
\newblock \url{http://dx.doi.org/10.1109/CVPR52688.2022.01553}.

\bibitem[Hebart et~al.(2023)Hebart, Contier, Teichmann, Rockter, Zheng, Kidder, Corriveau, Vaziri-Pashkam, and Baker]{contier2023}
Martin~N. Hebart, Oliver Contier, Lina Teichmann, Adam~H. Rockter, Charles Zheng, Alexis Kidder, Anna Corriveau, Maryam Vaziri-Pashkam, and Chris~I. Baker.
\newblock "things-meg", 2023.

\bibitem[Heilbron et~al.(2022)Heilbron, Armeni, Schoffelen, Hagoort, and De~Lange]{heilbron2022hierarchy}
Micha Heilbron, Kristijan Armeni, Jan-Mathijs Schoffelen, Peter Hagoort, and Floris~P De~Lange.
\newblock A hierarchy of linguistic predictions during natural language comprehension.
\newblock \emph{Proceedings of the National Academy of Sciences}, 119\penalty0 (32):\penalty0 e2201968119, 2022.

\bibitem[Khaligh-Razavi and Kriegeskorte(2014)]{KhalighRazavi2014}
Seyed-Mahdi Khaligh-Razavi and Nikolaus Kriegeskorte.
\newblock Deep supervised, but not unsupervised, models may explain it cortical representation.
\newblock \emph{PLoS Computational Biology}, 10\penalty0 (11):\penalty0 e1003915, November 2014.
\newblock ISSN 1553-7358.
\newblock \doi{10.1371/journal.pcbi.1003915}.
\newblock \url{http://dx.doi.org/10.1371/journal.pcbi.1003915}.

\bibitem[King et~al.(2026)King, Bel, Evanson, Gadonneix, Houhamdi, L{\'e}vy, Raugel, Santos~Revilla, Zhang, Bonnaire, Caucheteux, D{\'e}fossez, Desbordes, Diego-Sim{\'o}n, Khanna, Millet, Orhan, Panchavati, Ratouchniak, Thual, Brooks, Begany, Benchetrit, Careil, Banville, d'Ascoli, Dahan, and Rapin]{king2026neuralset}
J-R. King, C.~Bel, L.~Evanson, J.~Gadonneix, S.~Houhamdi, J.~L{\'e}vy, J.~Raugel, A.~Santos~Revilla, M.~Zhang, J.~Bonnaire, C.~Caucheteux, A.~D{\'e}fossez, T.~Desbordes, P.~Diego-Sim{\'o}n, S.~Khanna, J.~Millet, P.~Orhan, S.~Panchavati, A.~Ratouchniak, A.~Thual, T.~Brooks, K.~Begany, Y.~Benchetrit, M.~Careil, H.~Banville, S.~d'Ascoli, S.~Dahan, and J.~Rapin.
\newblock Neuralset: A high-performing python package for neuro-ai.
\newblock 2026.
\newblock \url{https://kingjr.github.io/files/neuralset.pdf}.
\newblock Preprint; URL will be updated when the paper lands on arXiv.

\bibitem[Lillicrap et~al.(2020)Lillicrap, Santoro, Marris, Akerman, and Hinton]{lillicrap2020backprop}
Timothy~P Lillicrap, Adam Santoro, Luke Marris, Colin~J Akerman, and Geoffrey~E Hinton.
\newblock Backpropagation and the brain.
\newblock \emph{Nature Reviews Neuroscience}, 21:\penalty0 335--346, 2020.

\bibitem[Liu et~al.(2022)Liu, Mao, Wu, Feichtenhofer, Darrell, and Xie]{ConvNext}
Zhuang Liu, Hanzi Mao, Chao-Yuan Wu, Christoph Feichtenhofer, Trevor Darrell, and Saining Xie.
\newblock A convnet for the 2020s.
\newblock In \emph{2022 IEEE/CVF Conference on Computer Vision and Pattern Recognition (CVPR)}, page 11966–11976. IEEE, 2022.
\newblock \doi{10.1109/cvpr52688.2022.01167}.
\newblock \url{http://dx.doi.org/10.1109/CVPR52688.2022.01167}.

\bibitem[Millet et~al.(2023)Millet, Caucheteux, Orhan, Boubenec, Gramfort, Dunbar, Pallier, and King]{millet2023realisticmodelspeechprocessing}
Juliette Millet, Charlotte Caucheteux, Pierre Orhan, Yves Boubenec, Alexandre Gramfort, Ewan Dunbar, Christophe Pallier, and Jean-Remi King.
\newblock Toward a realistic model of speech processing in the brain with self-supervised learning, 2023.
\newblock \url{https://arxiv.org/abs/2206.01685}.

\bibitem[Minderer et~al.(2022)Minderer, Gritsenko, Stone, Neumann, Weissenborn, Dosovitskiy, Mahendran, Arnab, Dehghani, Shen, Wang, Zhai, Kipf, and Houlsby]{Owl}
Matthias Minderer, Alexey Gritsenko, Austin Stone, Maxim Neumann, Dirk Weissenborn, Alexey Dosovitskiy, Aravindh Mahendran, Anurag Arnab, Mostafa Dehghani, Zhuoran Shen, Xiao Wang, Xiaohua Zhai, Thomas Kipf, and Neil Houlsby.
\newblock \emph{Simple Open-Vocabulary Object Detection}, page 728–755.
\newblock Springer Nature Switzerland, 2022.
\newblock ISBN 9783031200809.
\newblock \doi{10.1007/978-3-031-20080-9_42}.
\newblock \url{http://dx.doi.org/10.1007/978-3-031-20080-9_42}.

\bibitem[Naselaris et~al.(2011)Naselaris, Kay, Nishimoto, and Gallant]{naselaris2011encoding}
Thomas Naselaris, Kendrick~N Kay, Shinji Nishimoto, and Jack~L Gallant.
\newblock Encoding and decoding in {fMRI}.
\newblock \emph{Neuroimage}, 56\penalty0 (2):\penalty0 400--410, 2011.

\bibitem[Oord et~al.(2018)Oord, Li, and Vinyals]{oord2018representation}
Aaron van~den Oord, Yazhe Li, and Oriol Vinyals.
\newblock Representation learning with contrastive predictive coding.
\newblock \emph{arXiv preprint arXiv:1807.03748}, 2018.

\bibitem[Pedregosa et~al.(2011)Pedregosa, Varoquaux, Gramfort, Michel, Thirion, Grisel, Blondel, Prettenhofer, Weiss, Dubourg, Vanderplas, Passos, Cournapeau, Brucher, Perrot, and Duchesnay]{pedregosa2011scikit}
F.~Pedregosa, G.~Varoquaux, A.~Gramfort, V.~Michel, B.~Thirion, O.~Grisel, M.~Blondel, P.~Prettenhofer, R.~Weiss, V.~Dubourg, J.~Vanderplas, A.~Passos, D.~Cournapeau, M.~Brucher, M.~Perrot, and E.~Duchesnay.
\newblock Scikit-learn: Machine learning in {P}ython.
\newblock \emph{Journal of Machine Learning Research}, 12:\penalty0 2825--2830, 2011.

\bibitem[Radford et~al.(2021)Radford, Kim, Hallacy, Ramesh, Goh, Agarwal, Sastry, Askell, Mishkin, Clark, Krueger, and Sutskever]{Clip}
Alec Radford, Jong~Wook Kim, Chris Hallacy, Aditya Ramesh, Gabriel Goh, Sandhini Agarwal, Girish Sastry, Amanda Askell, Pamela Mishkin, Jack Clark, Gretchen Krueger, and Ilya Sutskever.
\newblock Learning transferable visual models from natural language supervision, 2021.
\newblock \url{https://arxiv.org/abs/2103.00020}.

\bibitem[Raugel et~al.(2025)Raugel, Szafraniec, Vo, Couprie, Labatut, Bojanowski, Wyart, and King]{DinoxBrain_hierarchy_forward_learning}
Joséphine Raugel, Marc Szafraniec, Huy~V. Vo, Camille Couprie, Patrick Labatut, Piotr Bojanowski, Valentin Wyart, and Jean-Rémi King.
\newblock Disentangling the factors of convergence between brains and computer vision models, 2025.
\newblock \url{https://arxiv.org/abs/2508.18226}.

\bibitem[Rumelhart et~al.(1986)Rumelhart, Hinton, and Williams]{Rumelhart1986}
David~E. Rumelhart, Geoffrey~E. Hinton, and Ronald~J. Williams.
\newblock Learning representations by back-propagating errors.
\newblock \emph{Nature}, 323\penalty0 (6088):\penalty0 533–536, October 1986.
\newblock ISSN 1476-4687.
\newblock \doi{10.1038/323533a0}.
\newblock \url{http://dx.doi.org/10.1038/323533a0}.

\bibitem[Schrimpf et~al.(2018)Schrimpf, Kubilius, Hong, Majaj, Rajalingham, Issa, Kar, Bashivan, Prescott-Roy, Geiger, et~al.]{schrimpf2018brain}
Martin Schrimpf, Jonas Kubilius, Ha~Hong, Najib~J Majaj, Rishi Rajalingham, Elias~B Issa, Kohitij Kar, Pouya Bashivan, Jonathan Prescott-Roy, Franziska Geiger, et~al.
\newblock Brain-score: Which artificial neural network for object recognition is most brain-like?
\newblock \emph{BioRxiv}, page 407007, 2018.

\bibitem[Siméoni et~al.(2025)Siméoni, Vo, Seitzer, Baldassarre, Oquab, Jose, Khalidov, Szafraniec, Yi, Ramamonjisoa, Massa, Haziza, Wehrstedt, Wang, Darcet, Moutakanni, Sentana, Roberts, Vedaldi, Tolan, Brandt, Couprie, Mairal, Jégou, Labatut, and Bojanowski]{dinov3}
Oriane Siméoni, Huy~V. Vo, Maximilian Seitzer, Federico Baldassarre, Maxime Oquab, Cijo Jose, Vasil Khalidov, Marc Szafraniec, Seungeun Yi, Michaël Ramamonjisoa, Francisco Massa, Daniel Haziza, Luca Wehrstedt, Jianyuan Wang, Timothée Darcet, Théo Moutakanni, Leonel Sentana, Claire Roberts, Andrea Vedaldi, Jamie Tolan, John Brandt, Camille Couprie, Julien Mairal, Hervé Jégou, Patrick Labatut, and Piotr Bojanowski.
\newblock Dinov3, 2025.
\newblock \url{https://arxiv.org/abs/2508.10104}.

\bibitem[Tschannen et~al.(2025)Tschannen, Gritsenko, Wang, Naeem, Alabdulmohsin, Parthasarathy, Evans, Beyer, Xia, Mustafa, Hénaff, Harmsen, Steiner, and Zhai]{Siglip2}
Michael Tschannen, Alexey Gritsenko, Xiao Wang, Muhammad~Ferjad Naeem, Ibrahim Alabdulmohsin, Nikhil Parthasarathy, Talfan Evans, Lucas Beyer, Ye~Xia, Basil Mustafa, Olivier Hénaff, Jeremiah Harmsen, Andreas Steiner, and Xiaohua Zhai.
\newblock Siglip 2: Multilingual vision-language encoders with improved semantic understanding, localization, and dense features, 2025.
\newblock \url{https://arxiv.org/abs/2502.14786}.

\bibitem[Wang et~al.(2023)Wang, Kay, Naselaris, Tarr, and Wehbe]{Wang2023}
Aria~Y. Wang, Kendrick Kay, Thomas Naselaris, Michael~J. Tarr, and Leila Wehbe.
\newblock Better models of human high-level visual cortex emerge from natural language supervision with a large and diverse dataset.
\newblock \emph{Nature Machine Intelligence}, 5\penalty0 (12):\penalty0 1415–1426, November 2023.
\newblock ISSN 2522-5839.
\newblock \doi{10.1038/s42256-023-00753-y}.
\newblock \url{http://dx.doi.org/10.1038/s42256-023-00753-y}.

\bibitem[Whittington and Bogacz(2017)]{Whittington2017}
James C.~R. Whittington and Rafal Bogacz.
\newblock An approximation of the error backpropagation algorithm in a predictive coding network with local hebbian synaptic plasticity.
\newblock \emph{Neural Computation}, 29\penalty0 (5):\penalty0 1229–1262, May 2017.
\newblock ISSN 1530-888X.
\newblock \doi{10.1162/neco_a_00949}.
\newblock \url{http://dx.doi.org/10.1162/NECO_a_00949}.

\bibitem[Yamins et~al.(2014)Yamins, Hong, Cadieu, Solomon, Seibert, and DiCarlo]{Yamins2014}
Daniel L.~K. Yamins, Ha~Hong, Charles~F. Cadieu, Ethan~A. Solomon, Darren Seibert, and James~J. DiCarlo.
\newblock Performance-optimized hierarchical models predict neural responses in higher visual cortex.
\newblock \emph{Proceedings of the National Academy of Sciences}, 111\penalty0 (23):\penalty0 8619–8624, May 2014.
\newblock ISSN 1091-6490.
\newblock \doi{10.1073/pnas.1403112111}.
\newblock \url{http://dx.doi.org/10.1073/pnas.1403112111}.

\bibitem[Zenke et~al.(2017)Zenke, Gerstner, and Ganguli]{zenke2017temporal}
Friedemann Zenke, Wulfram Gerstner, and Surya Ganguli.
\newblock The temporal paradox of hebbian learning and homeostatic plasticity.
\newblock \emph{Current opinion in neurobiology}, 43:\penalty0 166--176, 2017.

\bibitem[Zhuang et~al.(2021)Zhuang, Yan, Nayebi, Schrimpf, Frank, DiCarlo, and Yamins]{Zhuang2021}
Chengxu Zhuang, Siming Yan, Aran Nayebi, Martin Schrimpf, Michael~C. Frank, James~J. DiCarlo, and Daniel L.~K. Yamins.
\newblock Unsupervised neural network models of the ventral visual stream.
\newblock \emph{Proceedings of the National Academy of Sciences}, 118\penalty0 (3), January 2021.
\newblock ISSN 1091-6490.
\newblock \doi{10.1073/pnas.2014196118}.
\newblock \url{http://dx.doi.org/10.1073/pnas.2014196118}.

\end{thebibliography}
